\documentstyle[aaspp4]{article}  
\begin{document}
\title{\bf Environmental Dependence of the 
Fundamental Plane of Galaxy
Clusters}

\author{Christopher J. Miller\altaffilmark{1},
              Adrian L. Melott\altaffilmark{2,3},
                   and Patrick Gorman\altaffilmark{2}}

\altaffiltext{1}{Department of Physics \& Astronomy, University of Maine, Orono, ME 04469; chrism@perseus.umephy.maine.edu}
\altaffiltext{2}{Department of Physics \& Astronomy,
University of Kansas, Lawrence, KS 66045; melott@kusmos.phsx.ukans.edu; gorman@falcon.cc.ukans.edu}
\altaffiltext{3}{Department of Physics, Carnegie Mellon University, Pittsburgh, PA 15213}

 \begin{abstract} Galaxy clusters 
approximate a planar distribution
 in a three-dimensional parameter space 
which can be characterized by (for example)
 optical luminosity, half-light radius, and 
X-ray luminosity. 
We find the nearest neighbor
 clusters for those common to either of two  previous fundamental plane studies and 
a high-quality cluster redshift catalog. 
 Examining scatter about one plane in parameter space, we find 
a 2$\sigma$ result that it is
 dependent on nearest neighbor distance. 
Study of another sample of X-Ray clusters shows that those 
with high central gas densities are 
systematically (2.5$\sigma$) closer to neighbor clusters.
These results suggest a possible explanation for
recent evidence that X-ray cooling flow
clusters reside in crowded environments.

\end{abstract}


 \section{Introduction}

 Finding regularities in the characteristics 
of objects is one of the
 bridges from data collection to the 
development of theoretical
 understanding.  A now widely cited 
regularity concerns the Fundamental
 Plane (FP) of galaxies (Djorgovski and Davis 
1987; Dressler {\it et al.}
 1987).  When three parameters describing 
galaxies (various formulations
 are possible) are plotted, the points 
approximate a plane in the
 parameter space, telling us that  
there are really only two independent variables.

   More recently, the same approach has been 
applied to galaxy
   clusters and a Fundamental Plane has 
been found here, too
   (Schaffer {\it et al.} 1993; Adami {\it 
et al.} 1998).  Fritsch and
   Buchert (1999)-hereafter FB-examine not 
only the cluster FP but scatter
   about it. They define (non-uniquely) the FP in terms of 
total optical luminosity $L_o$, half-light 
radius $R_o$, and X-Ray luminosity $L_x$.  
FP can be used to predict and/or qualify 
physical characteristics. For
   instance, FB examine substructure in 
galaxy clusters, defined as a lack
   of symmetry and misalignment of 
concentric isophotes.  They find that
   clusters with strong substructure lie far 
from the FP.  They define the
   FP as the plane clusters take in the 
absence of substructure, and the
  \lq\lq empirical plane\rq\rq (EP) as the plane best 
fit to all clusters in their
   sample. EP and FP are very close together in their study. 

Fujita and Takahara (1999, hereafter FT) have also 
defined a different fundamental plane using other 
parameters:  gas density, core radius, and 
temperature as determined for a set of clusters published in
Mohr {\it et al.} (1999). The gas densities published in
Mohr {\it et al.} were determined differently for clusters
with and without cooling flows, $\rho_2$ and $\rho_1$ respectively.
When a cooling flow is present, FT convert $\rho_2$ to $\rho_1$ which
is more representative of a cluster's global structure. Thus, for
cooling flow clusters, $\rho_2$ represents the central gas density
as determined in the region of excess emission (the innermost region
of a cooling flow cluster) and
$\rho_1$ represents the converted gas density over a larger (but still central) region of the cluster
(see FT and Mohr {\it et al.} for further details).
FT use $\rho_1$ to create their fundamental plane which
has much less scatter 
than the plane of FB. 

Recently, it has proven possible to find 
regularities in the internal
   properties of clusters using a highly 
uniform, complete catalog of
   galaxy cluster redshifts (Miller {\it et 
al.} 1999a; Slinglend {\it et
   al.} 1998). Novikov {\it et al.} (1999) 
described an alignment between
   the wind direction  
distorting radio jets inside clusters
   and the long axis of the supercluster in 
which the cluster is embedded.
   Loken, Melott, and Miller (1999) present evidence 
that the existence of massive
   cooling flows is correlated with close 
proximity to other clusters.
The results presented here may present a partial
explanation for that result.

     \section{Data Analysis}

    We use a subset of the Abell (1958) and
Abell,Corwin, and Olowin (1989) cluster catalogs
with conservative cuts
    that enhance completeness.  Clusters are 
retained with richness
    $R\geq1$, redshift $z\leq0.10$, which 
are not close to the
    galactic plane, and have redshifts 
measured (not estimated) for
    multiple galaxies.  With these cuts, the 
remaining  clusters
    constitute a 98\% complete volume-
limited sample. Briefly, this
    sample has only minimal projection 
effects and few line-of-sight
    anisotropies (similar in degree to the 
APM cluster sample (Dalton {\it
    et al.} 1994)). In addition, most (80\%) of the 
clusters in our sample have three or more measured
    galaxy redshifts.
 Miller {\it et al.} (1999a), have shown
that cluster redshifts, determined from only
one galaxy, are in error by over 2500 km s$^{-1}$ 
at least 14\% of the time. Also, magnitude-redshift
relations typically have at least a 25\% scatter. 
A nearest-neighbor study, such as the one
presented here, requires accurate cluster
redshifts which can only come through multiple-galaxy observations.
   Additional evidence towards the quality of
this dataset come from
 the Voges et al. (1999) finding that
90\% of $R \ge 1$ Abell clusters (to $z = 0.09$)
are X-ray bright.
  Finally, we point out that the cluster
sample has a nearly constant number
density to $z = 0.10$. Therefore, even as additional
dimmer clusters are eventually observed, very few
will fall into the volume surveyed here.

\subsection{Environmental Correlations and the FB Plane}
       FB kindly provided data which was 
used in their study. 
We examined the scatter about their FP.  The FB
study used the logarithm of
       optical luminosity $L_o$, of X-ray 
luminosity $L_x$, and of
       half-light radius $R_o$ for the 
clusters; the plane is actually
in the space spanned by these axes.
(We refer to the axes which lie in the plane and orthogonal to it as the principal axes; the
measurables as the physical axes.)
Distances from the plane  are
       dimensionless and based on the 
$log_{10}$ of these measurables. This
       enabled us to associate a distance 
from the FP for each cluster. 
       We found 23 clusters present in FB 
for which we were able to
       reliably define a nearest neighbor 
distance. Clusters
       originally in FB but not in our study 
were excluded for not having
       a neighbor closer than the survey 
boundary or for being an $R=0$
       cluster.

We then looked for a correlation between
displacement from the fundamental plane
($d_p$) in parameter space and distance
from the nearest neighbor ($d_n$).
Displacement from the fundamental plane
was given a sign because the parameters
are not symmetric about it. There
are no published errors on the parameters
used to define the fundamental plane in FB.

Distances in redshift space to all of the clusters in our parent
sample were determined for a Friedman-Robertson-Walker
Universe with $q_0$ = 0 and
$H_0 = 100$ km s$^{-1}$ Mpc$^{-1}$.
(The choice of $H_0$ does not affect our correlations, and the effect of reasonable $q_0$ is much
smaller than other uncertainties which exist.)
To account for biasing effects caused
by the survey geometry, we have excluded
any cases where the distance to the edge of the
volume was smaller than $d_n$. This left
248 clusters with a nearest-neighbor distance.
When calculating $d_n$, we have
allowed for errors in each cluster's spatial coordinate
according to:
\begin{equation}
\triangle_i = \frac{7h^{-1}Mpc}{\sqrt{N_{cl}}}
\end{equation}
where $i = x,y, z$ and $N_{cl}$ is the number
of galaxies used for the mean cluster redshift.
We chose $7h^{-1}$Mpc since it is 
very near the typical velocity dispersion ($700$ km s$^{-1}$)
of rich clusters ({\it e.g.} Zabludoff {\it et al.} 1993).
We obtain $N_{cl}$ from a variety of sources
including Struble and Rood (1991), Postman,
Huchra and Geller (1992), Zabludoff {\it et al.} (1993),
Katgert {\it et al.} (1996) and Slinglend {\it et al.} 1998.
The error on each spatial coordinate is typically
around $1.5h^{-1}$Mpc,
which is rather conservative considering
an entire Abell radius is defined as $1.5h^{-1}$Mpc.
We then propagate through the errors on $x,y,z$ to 
calculate $\sigma_i$ for each nearest-neighbor distance, $d_n$.

We define a weighted correlation coefficient (Bevington, 1969)
\begin{equation}
K \equiv \frac{s_{d_n, d_p}^2}{s_{d_n}s_{d_p}}
\end{equation}
where
\begin{equation}
s_{d_n, d_p}^2 \equiv
\frac{\frac{1}{N-1}\sum[\frac{1}{{\sigma_i}^2}(d^i_n-\bar{d}_n)(d^i_p -
\bar{d}_p)]}{\frac{1}{N}\sum{\frac{1}{\sigma_i}^2}}
\end{equation}
and
\begin{equation}
s_{d_n}^2 \equiv
\frac{\frac{1}{N-1}\sum[\frac{1}{{\sigma_i}^2}(d^i_n-\bar{d}_n)^2]}{\frac{1}{N}\sum\frac{1}{{\sigma_i}^2}}
\end{equation}
and
\begin{equation}
s_{d_p}^2 \equiv  \frac{1}{N-1}\sum(d^i_p-\bar{d}_p)^2.
\end{equation}

       We find $K=-0.41$, a fairly strong 
correlation.  This is a 2.0$\sigma$
       result indicating 97.6\% confidence 
that it did not arise as a
       chance fluctuation.  
Ignoring the sign of $d_p$ considerably weakens the result.
If we apply equal weights, the correlation increases
to $K = -0.53$ or a 2.5$\sigma$ result.

We also looked for correlations between $d_n$ and the three physical axes
as well as the other two principal axes
of the FP. To calculate $K$ in these cases, we replaced
$d_p$ in equations 3,4,5 with the parameters under examination. 
We found no significant correlations between $d_n$ and any of the other
quantities.
Our main finding -- that the distance a cluster lies from
the fundamental plane (defined by FB) in parameter space
correlates with it's nearest neighbor distance -- cannot
be attributed to any single parameter alone.
Nor do we find evidence that the the position projected onto the FP
is dependent upon $d_n$.

\subsection{Environmental Correlations and the FT Plane}

FT examined and systematized data gathered by 
Mohr {\it et al.} (1999).  We also examined 
the FT data, using $\rho_1$, core radius, and gas temperature;
there were 14 clusters with
nearest neighbors in our data set. 
In this case, $d_p$ has a rather 
small range, since the FT plane is thin, and we found the reasonable 
result that there were no significant 
correlations between it and $d_n$, nor are there 
between $d_n$ and the other two principal axes of this ribbon-like FP.
Neither was $d_n$ signficantly correlated with core radius, temperature 
or converted gas densities (all $\rho_1$ as in FT) considered alone.
However, when we examine $d_n$ vs. central gas density --using
$\rho_2$ when a cooling flow is present and otherwise $\rho_1$--
we find a
2.5{$\sigma$} (99.4 \% confidence) result - 
a correlation coefficient of $K=-0.67$. 
In this case, we used errors on $d_n$ as described earlier and
errors on $\rho$ as published in Mohr {\it et al.}. If we apply
equal weights to each value for $d_n$ and $\rho$, the correlation
falls to $K = -0.55$.
This property of X-ray 
clusters -- high central gas density in 
clusters close to other clusters -- may explain a key result of 
Loken {\it et al.} (1999):  the tendency of 
a cooling flows to occur in clusters with near 
neighbors.  It appears likely that these 
crowded clusters have gas densities high 
enough to give the prerequisite short 
cooling times.

\subsection{Selection Effects}
Although the Abell/ACO sample used to
define nearest neighbors is
the most complete of its kind
available, there remains the
possibility that there are undetermined
optical selection effects. For instance,
clusters in regions of low galactic
neutral hydrogen density (n$_H$) might
appear brighter. In addition, in regions
of low dust and n$_H$,  Abell (1958) may have
been more likely to find a cluster
neighbor in close proximity. 
However, neither 
$L_o$ or $R_o$
alone shows a strong correlation towards $d_n$.
If it did, we might suspect it were a result of superposition
or a similar artifact.

With the recent construction of the Schlegel {\it et al.} (1998)
reddening maps, it is possible to determine whether our
highest $L_o$ clusters happen to lie in abnormally low
regions of galactic extinction. A strong anti-correlation
between regions of low HI column density and richer Abell
clusters has been found by Nichol \& Connolly (1996). Such
an effect could result in a greater number of close cluster
pairs, which are also brighter, to contaminate our small sample.
Therefore, we have compared the magnitude of reddening, E(B-V),
as determined
from the Schlegel {\it et al.} map for $\sim 4000$ random locations
on the sky to those centered on our 23 clusters. A Kolmogorov-Smirnov (K-S) test shows
that the HI distributions are nearly identical and we are not
sampling regions of abnormally low HI column densities.
In addition, we
find no correlations between E(B-V) and $L_o$, $d_n$ or $d_p$.

Another possible selection effect that cannot explain our results
is richness. Our sample of 23 clusters contains mostly (18/23)
$R=1$ clusters and the remaining are $R = 2$. The mean nearest
neighbor distances as a function of richness for the entire
complete catalog are  R=1: 19.2, R=2: 16.4, and R=3: 22.3 (in
$h^{-1}$Mpc). Furthermore, clusters show no evidence of
richness-dependence in terms of their displacement from FP.

\section{Conclusions}

Cluster properties depend on their environment as parameterized by
the distance to their nearest external cluster. After ruling out
a substantial number of possible selection effects, we find here that
there are strong correlations between cluster properties and the 
proximity of other clusters.

 From the FB study, we find that clusters far from and \lq\lq below\rq\rq~ the FP tend to be 
isolated.  As we move \lq\lq up\rq\rq~ toward their plane in the direction of $L_o^{0.81} 
R_o^{-0.84} L_x^{-0.21}$, the clusters tend to be much closer in physical space 
to other clusters.  This suggests optically brighter, more compact clusters are in more 
crowded environments.

 From FT, we did not find any significant correlations along their principal 
axes.  However, there was a strong tendency for central gas density to be 
higher in X-ray clusters which are close to other clusters.  This is
consistent with the previous
paragraph, and may explain the propensity of such crowded clusters to initiate 
cooling flows (Loken {\it et al.} 1999).

We know from the FB analysis also that clusters close to their FP have much less 
substructure.  Putting this together with our results we can summarize as 
follows: Clusters in crowded environments tend to have less substructure and higher central 
gas densities.  Together this provides an explanation for the Loken {\it et al.} 
(1999) result:  a relaxed cluster with little substructure provides the 
symmetry and high central density needed to set up a massive cooling flow, and 
these conditions are found in those clusters located in close proximity with 
other clusters.

This is reasonable on theoretical grounds: perturbations of a given mass scale
(in this case clusters) which lie in a larger region of high amplitude are more
likely themselves to be of high amplitude.  A higher amplitude implies earlier
collapse and more time to relax.  Such relaxed clusters are more likely to take
a more \lq\lq universal\rq\rq~ locus in parameter space, with less substructure (artifact
of initial conditions including merger history) a higher central gas density
and the ability to initiate a cooling flow.  This picture fits together
the cooling flow results of Loken et al. (1999), substructure correlations
found by FB, and the correlations we found by environmental study of the
FB and FT data groups.

Although our sample sizes (23 and 14) are rather small, results of fairly high 
confidence exist.  We take this as evidence of the strength of the effect 
combined with the superior characteristics of the redshift catalog after the 
cuts were taken.

       \noindent

       {\bf Acknowledgments}

  CM was 
funded in part by the National
       Aeronautics and Space Administration and the Maine Science and
       Technology Foundation.
       ALM acknowledges the support of the NSF-EPSCoR program, 
       the hospitality of Carnegie Mellon University during part of this work,
       and authors FB for sharing their data. Jim Fry and David Batuski
       made helpful suggestions.

\begin{figure*}
\plotone{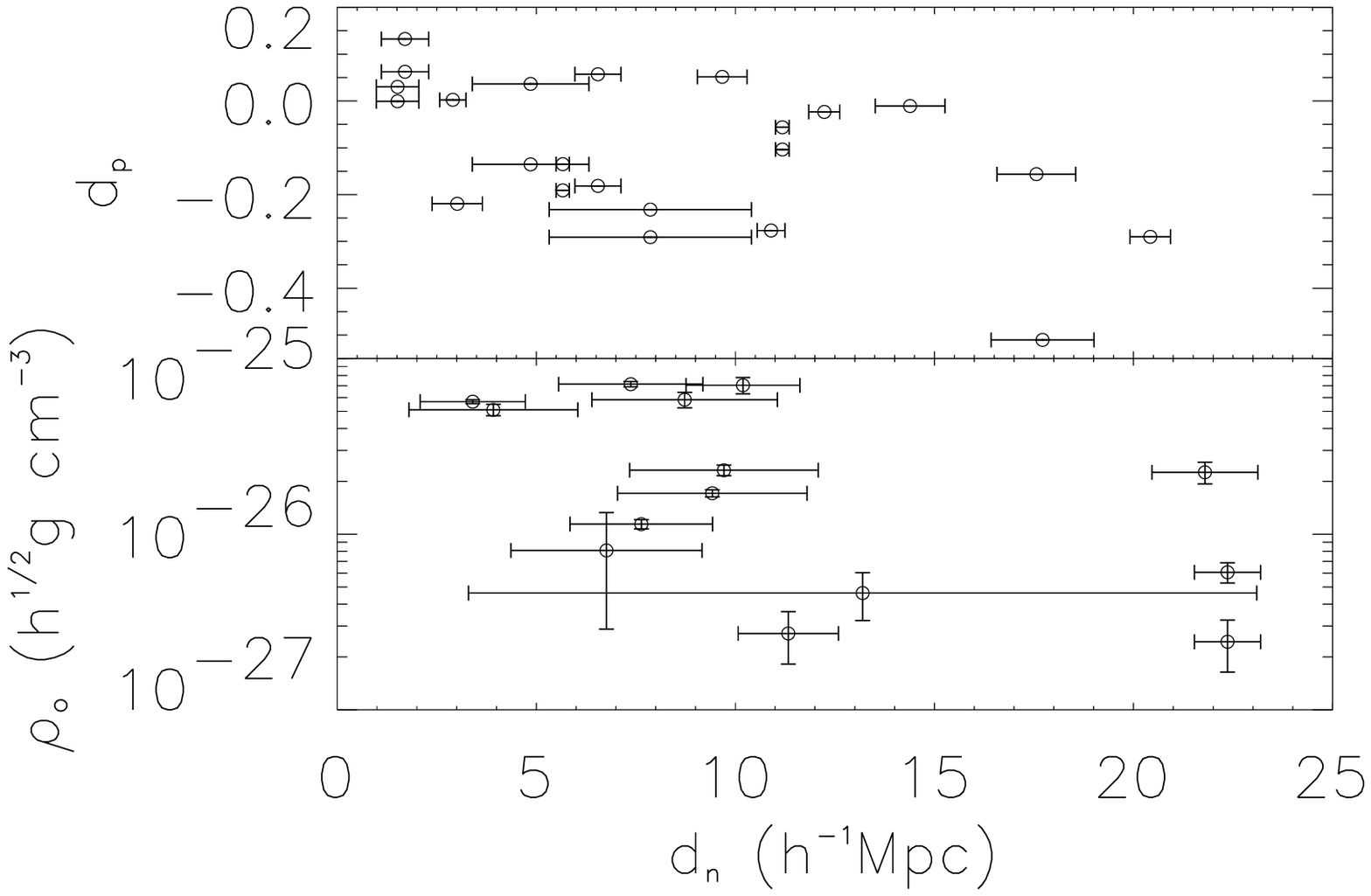}
\caption[]{
In both panels, the x-axis is the distance to the nearest-neighbor galaxy
cluster to the given cluster within our sample.  The top panel shows the
distance to the FP in the FB paper in dimensionless (log) units.  The
bottom panel shows the X-ray central gas density, $\rho_o$, as
published in Mohr {\it et al.} 1999.}
\end{figure*}

\end{document}